# THz emission from a $Bi_2Sr_2CaCu_2O_{8+\delta}$ cross-whisker junction

Yoshito Saito[1,2*], Shintaro Adachi[1], Ryo Matsumoto[3], Masanori Nagao[4], Shuma Fujita[5],

Ken Hayama[5], Kensei Terashima[1], Hiroyuki Takeya[1], Itsuhiro Kakeya[5],

and Yoshihiko Takano[1,2]

[1]*MANA, National Institute for Materials Science (NIMS), 1-2-1 Sengen, Tsukuba, Ibaraki 305-0047, Japan*

[2]*Graduate School of Pure and Applied Sciences, University of Tsukuba, 1-1-1 Tennodai, Tsukuba, Ibaraki 305-8577, Japan*

[3]*ICYS, National Institute for Materials Science (NIMS), 1-2-1 Sengen, Tsukuba, Ibaraki 305-0047, Japan*

[4]*Center for Crystal Science and Technology, University of Yamanashi, 7-32 Miyamae, Kofu, Yamanashi 400-0021, Japan*

[5]*Department of Electronic Science and Engineering. Kyoto University, Kyoto daigaku-katsura, Nishikyo-ku, Kyoto 615-8510, Japan*

E-mail: SAITO.Yoshito@nims.go.jp

Cuprate superconductor $Bi_2Sr_2CaCu_2O_{8+\delta}$ (BSCCO) has been a promising candidate of a coherent, continuous, and compact THz light source owing to its intrinsic Josephson junction inside the crystal structure. In this paper, we utilized BSCCO cross-whisker junctions to produce THz emitter device using the whisker crystals which can be easily obtained compared with single crystals. As a result, we have successfully observed the emission from the cross-whisker intrinsic Josephson junction with frequency of ~0.7 THz, which is a first observation of THz emission from whiskers to our knowledge. Our findings would enlarge the applicability of BSCCO superconductors for the THz emission source.

The terahertz (THz) electromagnetic wave has been attracting interests in various application uses [1,2], such as less-invasive imaging of organic materials and fingerprinting chemical substances. However, these applications have been hindered due to the absence of practical technologies to generate THz waves, called "THz gap". From the discovery of the Josephson effect [3], the use of superconductors with Josephson junctions has been studied as a high-frequency electromagnetic wave source. After the discovery of cuprate superconductors, an innovative finding has been brought by Kleiner *et al.*, who discovered that Josephson junctions naturally exist in the crystal structure of $Bi_2Sr_2CaCu_2O_{8+\delta}$ (BSCCO) itself [4], called intrinsic Josephson junction (IJJ). Theoretical calculations predicted the emission of electromagnetic wave with THz frequency from IJJ of cuprate superconductors under magnetic field and DC current bias [5]. Such a predicted intense THz emission has been confirmed experimentally, by forming a mesa structure in BSCCO single crystal [6]. The mesa structure is expected to work as a cavity resonator, therefore the cavity resonance frequency of the system is governed by the dimensions of a mesa structure.

To date, all these superconducting THz emitters are made of high quality BSCCO single crystals grown by the traveling-solvent floating zone method. On the other hand, BSCCO can also be grown in the form of "whisker" crystal [7,8] that grows in needle shape. These BSCCO whiskers can be grown with a conventional electric furnace, and the growth process is far easier than that of the TSFZ method but holds quite high crystallinity. However, the THz emission from BSCCO whiskers has not been reported so far. This would be mainly because the width of available whiskers is smaller than that of mesas in single crystals, not being able to fulfill the cavity resonance condition. Therefore, a new method should be developed for generation of THz waves from BSCCO whiskers.

Our approach to solve the above problem has two essential points. (i) The use of the so-called "cross-whisker (CW) junction" [9-11], which is micro-fabrication free intrinsic Josephson junctions, made of two BSCCO whiskers put on a substrate crosswise and bonded together by heat treatment to form a homogenous joint between whiskers. (ii) Fabrication of two notches at the distance of several tens of μm on the lower whisker by focused ion beam, whose depth is ~1 μm. This increases the number of corresponding IJJs participating for THz emission. Hereafter, we call this stack of IJJs fabricated in a cross-whisker junction as CW-IJJ.

In this paper, we introduce a fabrication of a CW-IJJ using BSCCO whiskers as a THz emitter device. The emission of the THz wave has been successfully observed, which is a first report of THz emission from whiskers to our knowledge. Our study opens a possibility of the CW-IJJ being a unique THz source, such as an antenna-like arrangement with enhanced directivity or taking advantage of interference between multiple whiskers, due to its capability of taking various configurations.

BSCCO whisker crystals were grown by Te-doped method [8]. High purity powder of $Bi_2O_3$, $SrCO_3$, $CaCO_3$, CuO and $TeO_2$ were weighed at the cation ratios of Bi : Sr : Ca : Cu : Te = 2.1 : 1.9 : 2.0 : 2.0 : 0.5 as the starting materials. Precursor powders were synthesized by solid-state reaction through grinding and calcination for three times. The calcination temperatures were 760 °C for the first calcination, then, the second and third calcination were done at 790 °C and 820 °C, respectively. After calcinations, precursor powders were pressed into pellets (~0.5 g), which was further annealed at 840 °C for 48 hours under oxygen flow to obtain BSCCO whiskers. The composition ratio of an as-grown whisker was Bi : Sr : Ca : Cu = 2.39 : 1.58 : 1.37 : 2.00, which is analyzed by energy dispersive x-ray spectroscopy.

A CWJ was fabricated in the following procedures. First, the cleaved surfaces of BSCCO whiskers were attached each other crosswise at the angle of 90°on an MgO (100) substrate. Next, they were annealed at the temperature of 850 °C for 30 minutes under oxygen flow. After that, silver paste was applied to four ends of whiskers to form terminals for four probe measurements of resistance and current-voltage characteristics.

To increase the number of corresponding IJJs in the fabricated cross-whisker junction device, we exposed the lower whisker to focused ion beam for producing shallow notches with the depth of ~1 μm. The distance between two notches was set to be 85 μm, typical value of resonant length in existing THz emitting mesas. Fig. 1 shows a scanning ion microscope image of the fabricated CW-IJJ device with a constructed electrical circuit for this study. The tips of each whiskers were connected to either a function generator or a multimeter. Thus, four-terminal measurement is possible with the CW-IJJ.

For THz measurements, the device was mounted on a cold plate of a pulse tube cryocooler with quartz window (V208PLS, DAIKIN INDUSTRIES). The THz emission was detected using a liquid He cooled InSb hot electron bolometer with $120 cm^{-1}$ low-pass filters (QFI/2BI, QMC Instruments). The bolometer signal was obtained by a lock-in amplifier (5610B, NF Electronic Instruments) with a modulation frequency of 23.5 Hz by an optical chopper. The spectra of the emission were obtained by an FT-IR spectrometer (FARIS-1, JASCO). First, the emission was directly detected by the bolometer. Then the emission frequency was measured by FT-IR spectrometer with the same bolometer.

Figure 2 (a) shows the temperature dependence of the resistance ($R$) of the CW-IJJ. The resistance decreased linearly from the room temperature, then an up-turn appears around 150

K, this is consistent with the previous reports insisting that as-grown whiskers are in over-doped state [12]. Then a small drop of the resistance was observed at 110 K, which is due to the presence of $Bi_2Sr_2Ca_2Cu_3O_{10+\delta}$ intergrowth component. The onset of superconducting transition temperature was estimated to be 79 K while the device showed zero-resistance at 72 K. The four-terminals attached to our CW-IJJ allowed us to confirm zero resistance of the sample, unlike previous THz emitting mesas with two-terminal method.

Figure 2 (b) shows the current-voltage (*I-V*) characteristics of the sample for several temperatures at 40, 45, 50, 55 and 60 K. As we increase the voltage, the *I-V* curves show plateau at I~15 mA and then the current increases at higher voltages, indicating that all the IJJs in the sample are in the normal state above the critical current of ~15 mA in this temperature range. Next, as we decrease the voltage from this normal state, the current shows a monotonic decrease with hysteresis.

To detect the electromagnetic emission from the device, we placed the hot electron bolometer and focusing lens system in front of the CW-IJJ device. Figure. 3 shows a *I-V* characteristic at $T$ = 50 K and bolometer output signal recorded simultaneously. The bolometer signal showed clear peak structures in the return current sweep process, therefore we have successfully observed the emission from our CW-IJJ at $I \sim 10$ mA and $V \sim 0.7$ V, corresponding to so-called “Low Bias” regime. We have observed the emission in the temperature range of 40 to 60 K. The sample showed the maximum emission intensity at 50K.

We also evaluated the emission frequency of the device. For this purpose, the sample signal was recorded in the FT-IR spectrometer setup. Figure 4 (a) is an enlarged view of a *I-V* characteristic of the THz emitting region at $T$ = 50 K and the inset shows the whole picture of the *I-V* characteristic. We have obtained the emission spectra at three points in the same branch as shown in Fig. 4 (a). Figure 4 (b) shows the emission spectrum for each bias points with the fitting curves of the peaks using Lorentzian function. The emission frequency decreases with reducing the bias voltage. The number ($N$) of IJJs contributing to the THz emission can be estimated in the formula of AC Josephson effect as follows : $f = 483.6 \times V / N$, where $f$ corresponds to the emission frequency [GHz], $V$ is voltage across the all IJJs [mV]. In the current case, estimated $N$ is 431, which corresponds to the effective height of the IJJ stack as 0.65 μm.

To our knowledge, this is a first observation of the THz emission generated from BSCCO whiskers. Compared with bulk single crystals, the whiskers can be easily grown by simple furnace. Moreover, the cross-whisker IJJ device may provide us the possibility for extended applications of THz emitters, as it consists of multiple superconductors and there is a fair room of the possible combinations and spatial configurations. For instance, addition of several lower whiskers might generate highly directed THz emission as expected from its similarity with Yagi-Uda antenna array.

In conclusion, we fabricated a BSCCO whiskers-based CW-IJJ device as a THz emitter source. We have successfully observed an emission of electromagnetic waves from the device at the frequency of ~0.7 THz in the temperature range between 40 to 60 K, which would be the first report of a THz emission from over-doped BSCCO crystal to our knowledge. Our findings indicate that, in addition to conventional single crystals, BSCCO whiskers are also capable of generating THz emission.

**Acknowledgments**

We are grateful to S. Harada, T. Ishiyama, and S. Iwasaki for all the help. This work was supported by JSPS KAKENHI Grant Numbers 20H05644, 19H02177, 18K04717, and JST-Mirai Program Grant Number JPMJMI17A2.

## Figure Captions

**Fig. 1.** A scanning ion microscope image of the CW-IJJ and a schematic of the associated circuit.

**Fig. 2.** (a) Temperature dependence of the resistance for the CW-IJJ. (b) Current-Voltage characteristics of the CW-IJJ at several temperatures.

**Fig. 3.** Current-Voltage characteristics and hot electron bolometer signal response for the CW-IJJ at T = 50 K.

**Fig. 4.** (a) An enlarged I-V characteristic of the CW-IJJ at T = 50 K with bias points A-C, where FT-IR spectroscopy was measured. The inset shows the whole I-V characteristic. (b) The spectra taken at bias points A-C with Lorentzian fitting curves.

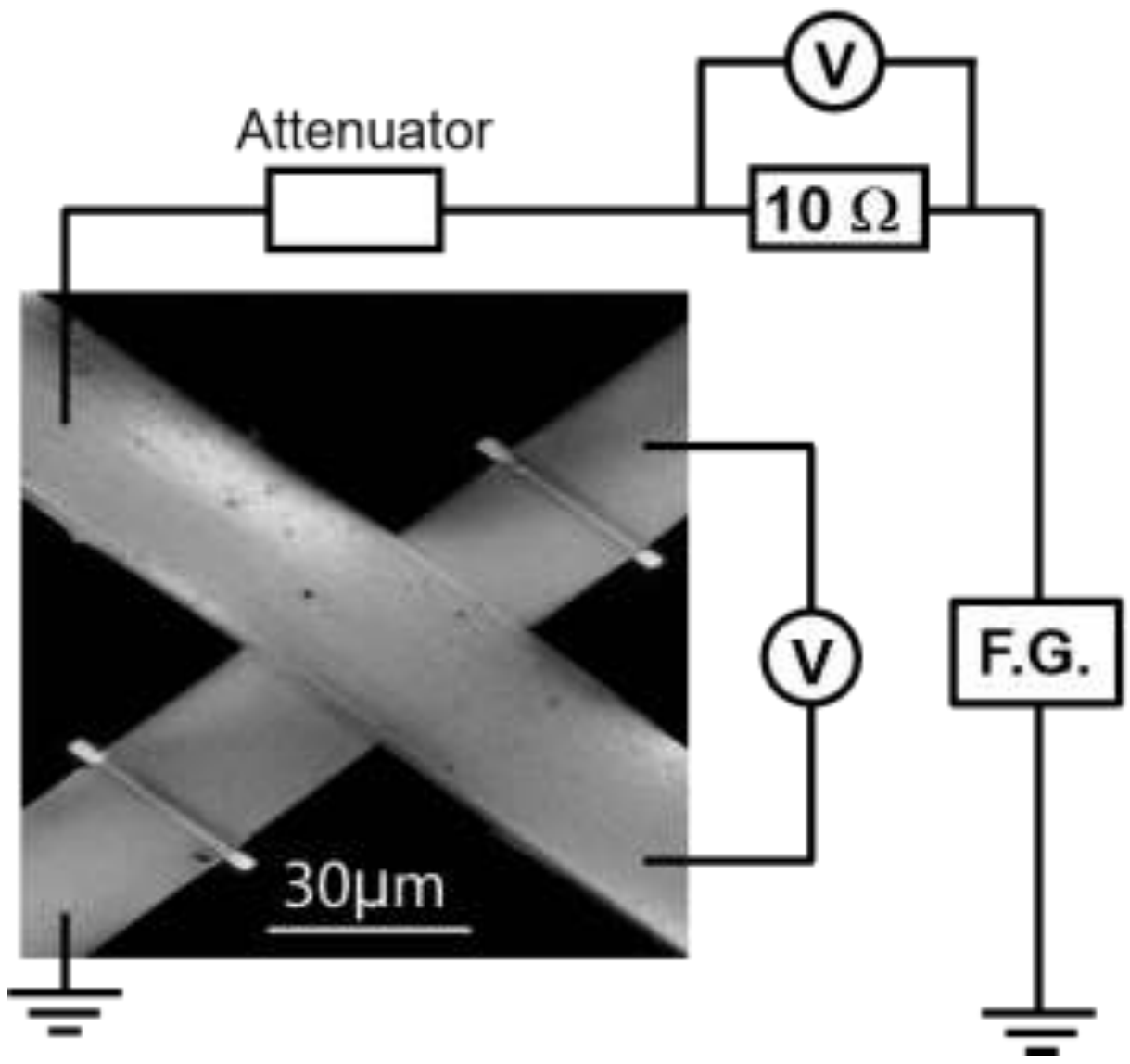

Attenuator
V
10 Ω
V
F.G.
30μm

Fig. 1.

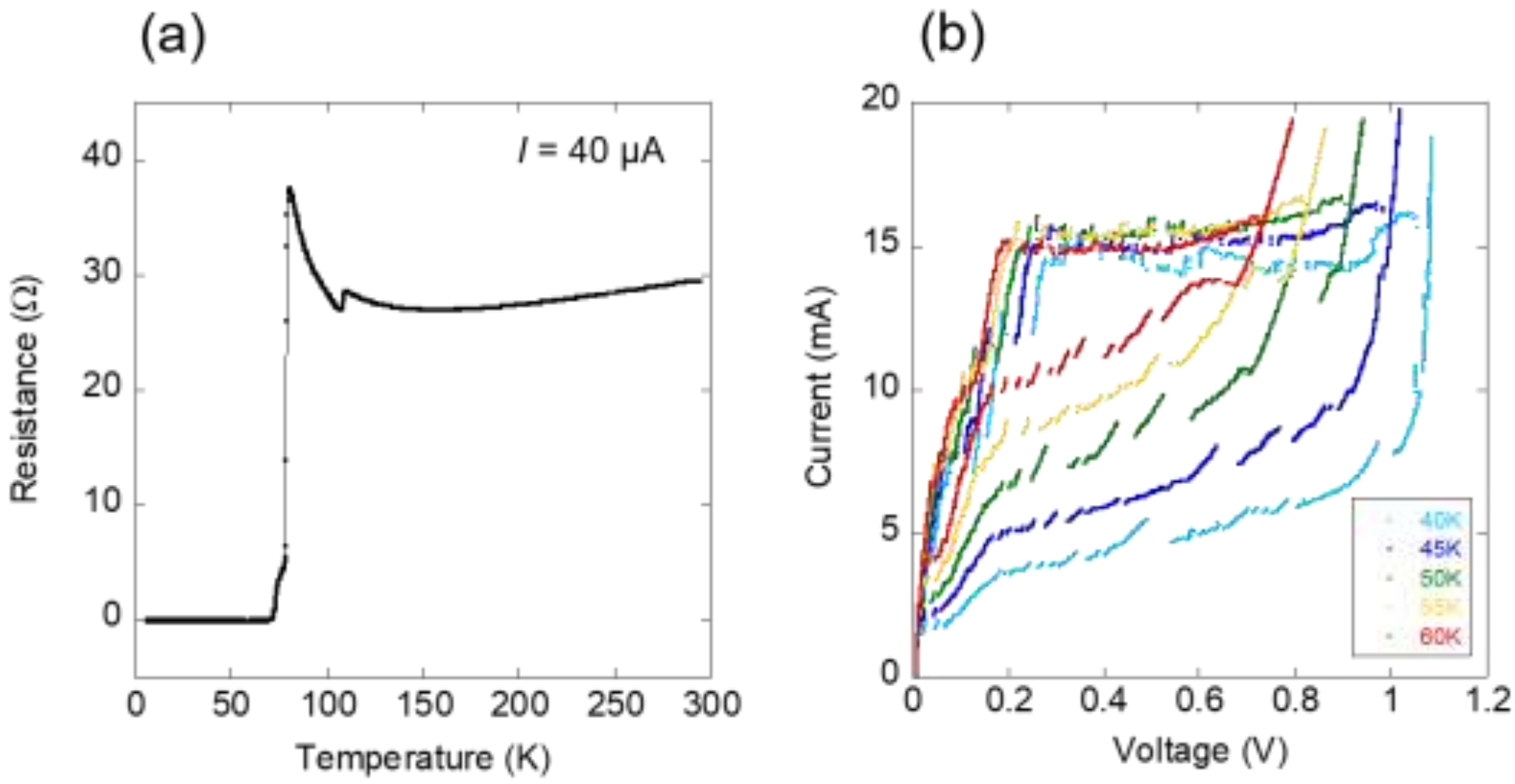

(a)
I = 40 μA
Resistance (Ω)
Temperature (K)
(b)
Current (mA)
Voltage (V)
40K
45K
50K
55K
60K

Fig. 2.

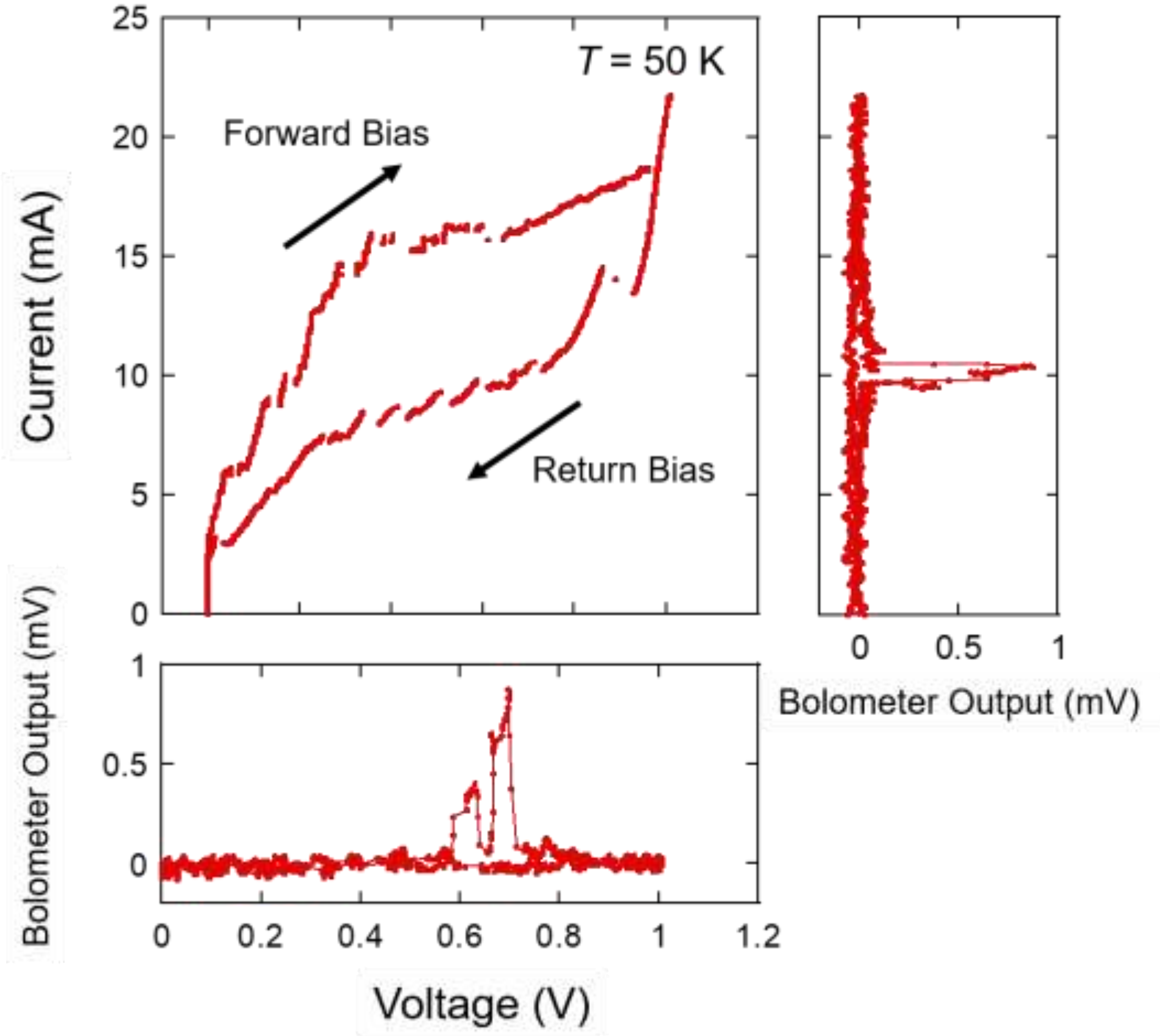
T = 50 K
Forward Bias
Return Bias
Current (mA)
Bolometer Output (mV)
Voltage (V)
Bolometer Output (mV)
25
20
15
10
5
0
1
0.5
0
0
0.2
0.4
0.6
0.8
1
1.2
0
0.5
1

Fig. 3.

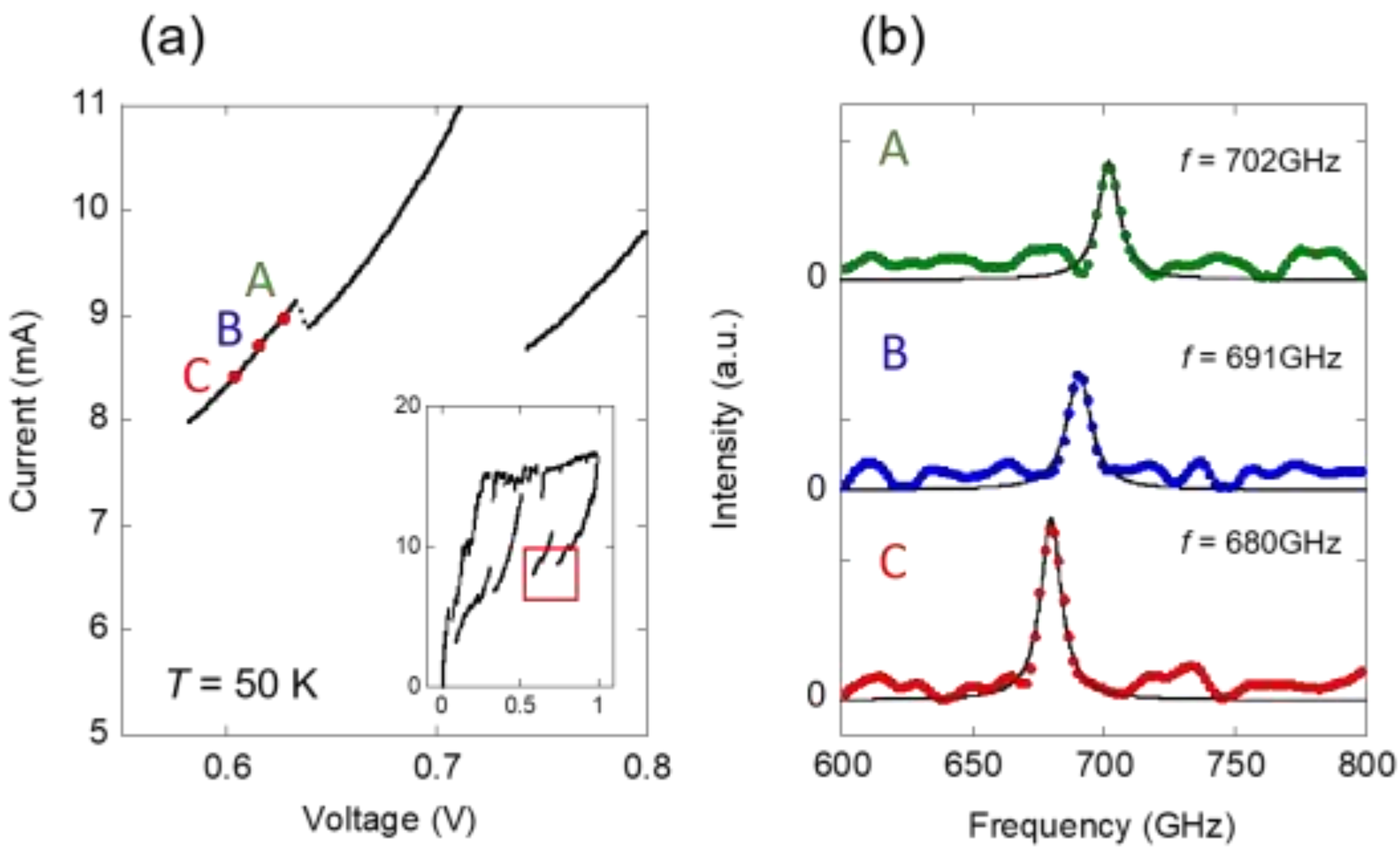

(a)
(b)
Current (mA)
Voltage (V)
T = 50 K
A
B
C
Intensity (a.u.)
Frequency (GHz)
f = 702GHz
f = 691GHz
f = 680GHz

Fig. 4.